\documentstyle[preprint,aps,version2]{revtex}

\renewcommand{\baselinestretch}{1.5}
\renewcommand{\arraystretch}{1.5}
\begin{document}

\vspace{3mm}
\renewcommand{\baselinestretch}{2}
\small \normalsize
\hspace{6cm} Subm. to Phys.Rev.Lett.; November 9, 1993

\begin{center}
 { \large A Simple Formula for the Persistent Current in\\
       Disordered 1D Rings: Parity and Interaction Effects}\\
 \end{center}
\vspace{3mm}
\begin{center}
 {\bf Alexander O.Gogolin$^{\dag}$ }\\
 {\em Landau Institute for Theoretical Physics,
  Russian Academy of  Sciences,} \\
 { \em Kosygina str.2, Moscow, 117940, Russia}\\
{\bf Nikolai V. Prokof'ev$^{\dag \dag }$}\\
{\em Physics Department, University of
 British Columbia, 6224 agricultural RD.,} \\
{\em Vancouver, BC, Canada V6T 1Z1}\\
\end{center}
\vspace{2mm}

\renewcommand{\baselinestretch}{2}
\small \normalsize

\begin{abstract}
The general expression for the persistent current of 1D noninteracting
electrons in a disorder potential with smooth scattering data is derived
for zero temperature.
On the basis of this expression the parity effects are discussed.
It is shown that the electron-electron interaction
may give rise to an unusual size-dependence of the current amplitude
due to  renormalized backscattering from the disorder potential.
\end{abstract}

\begin{center}
{$^{\dag}$ present address: Institute Laue-Langevin, B.P.156,
  38042 Grenoble, Cedex 9, France; Email: gogolin@gaston.ill.fr}\\
{$^{\dag \dag }$ permanent address:  Kurchatov Institute,
 Moscow, 123182, Russia}
\end{center}

\newpage

The persistent current in 1D mesoscopic rings,
being a manifestation of the Aharonov-Bohm effect for many-body systems,
addresses physical questions of principle.
Although such an effect was predicted several decades ago\cite{PP},
in normal state systems it was only recently discovered experimentally.
The experiments were performed on ensembles of many mesoscopic rings\cite{MRE}
as well as on single (or several) rings \cite{SRE}.
Generally, the theory predicts a magnitude of the current, which is much
less then one experimentally observed\cite{RT},
although there are some cases for which the theory and experiment
seem to agree\cite{NR}.
To the best of our knowledge there are no such studies  of the
interplay between the  disorder and electron-electron interaction in
1D which would
take into account nonperturbative effects in the persistent current
(on the other hand, several numerical
investigations have been curried out\cite{NR}).

The aim of this paper is to derive a simple formula for the persistent
current of noninteracting electrons in a single-channel mesoscopic ring
with a disorder potential and  to study the interplay between the
disorder and electron-electron interaction in 1D.

For $N$ spinless electrons on the ring of the length $L$ under
Aharonov-Bohm conditions the persistent current is given by the formula:
\begin{eqnarray}
&&j(\varphi )=j_{0}(\varphi )+j^{par}(\varphi )+{\large O}(1/L^{2});
\nonumber \\
&&j_{0}(\varphi )=\frac{v_{_F}}{\pi L}
\frac{\sqrt{{\em T}_{F}}\sin \varphi}{\sqrt{1-{\em T}_{F} \cos^{2} \varphi}}
F(p_{F},\varphi ); \;\;\; F(p,\varphi )=
\cos^{-1} \left( \sqrt{{\em T}(p)}\cos \varphi \right) , \\
&&j^{par}(\varphi )=-\frac{v_{_F}}{L}
\frac{\sqrt{{\em T}_{F}}\sin \varphi}{\sqrt{1-{\em T}_{F} \cos^{2} \varphi}},
\;\;\; (N=even); \;\;\;\;\;\;\; j^{par}(\varphi )=0,
\;\;\; (N= odd)\;,\nonumber
\end{eqnarray}
where $\varphi =2\pi\Phi /\Phi_{0}$ ($\Phi$ is the magnetic flux
passing through the ring) and ${\em T_{_F}}\equiv {\em T}(p{_F}) $
 is the transmission
coefficient of the ring at the Fermi energy.
The above formula explicitly shows that the persistent current
is a purely Fermi surface effect.
Moreover, it is parametrized by a single number - ${\em T_{_F}}$ (i.e.
the only characteristic of the potential,
which determines the flux dependence of the current, is ${\em T_{_F}}$)
\cite{LF}.

{\bf \em Derivation}. The single electron Schr\"{o}dinger equation
\begin{equation}
\left\{ \varepsilon_{0} \left( -i\partial_{x}
\right) +V(x)
-\varepsilon \right\} \psi =0,
\label{2}
\end{equation}
where $V(x)$ is the disorder potential and
$\varepsilon_{0}(p)$ is the dispersion law,
should be solved under the twisted boundary condition:
$\psi (x+L)=e^{i\varphi}\psi (x)$.
It was recognized already in the pioneer papers \cite{PP}
that in this Bloch functions problem
 the whole ring plays the role of the elementary cell
and $\varphi$ - the role of the quasimomentum.
The ground state energy as a function of the flux
is given by the sum
$E_{0}(\varphi )=\sum\varepsilon_{\lambda}(\varphi )$
over $N$ lowest values of the band spectrum at fixed $\varphi$.

Suppose, for clarity, that the potential $V(x)$ is localized
within the region of a radius $a<L$
(actually, the relation between $a$ and $L$ will be shown
to be unimportant).
The wave-function $\psi (x)$ can be then written in the form
$\psi (x)=Ae^{ipx}+Be^{-ipx}$ from the left of the potential
and in the form $\psi (x)=Ce^{ipx}+De^{-ipx}$ from the right of it.
The coefficients $C$ and $D$ can be expressed through $A$ and $B$
by making use of the scattering data of $V(x)$ (transfer matrix).
The twisted boundary condition  gives then
(after elementary algebra) the equations for the spectrum:
\begin{equation}
pL =  \Phi_{+}(p,\varphi ) \;\;\;( n=0);\;\;\;\;\;
pL = 2\pi n + \Phi_{\pm}(p,\varphi )\;\;\; ( n=1,2,...);  \\
\label{5}
\end{equation}
where $\Phi_{\pm}(p,\varphi )=\delta (p) \pm F(p,\varphi )$ and
$\delta (p)$ is the forward scattering phase
defined by the
scattering solution: if $A=1$ and $D=0$, then $C=\sqrt{{\em T} (p)}
e^{i\delta (p)}$.

To make progress with the Eq.(\ref{5}), the idea is to expand the
solution in $1/L$.
This is simplest when $a<<L$ (e.g. $V(x)$ represents a single
scatterer) and $\varepsilon_{0}(p)=p^{2}/2m$. Then one can write:
\begin{equation}
p_{n}=k_{n}+\frac{1}{L} \Phi_{\pm}(k_{n})+
\frac{1}{L^{2}} \Phi_{\pm}(k_{n})
\frac{\partial\Phi_{\pm}(k_{n})}{\partial k}
+{\large O}\left( \frac{1}{L^{3}}\right) ;\;\;\; \; k_{n}=2\pi n/L
\label{6}
\end{equation}
For the ground state energy we therefore have:
\begin{equation}
\left.
E_{0}=\frac{1}{m}\sum_{n}\left\{ k^{2}+\frac{2}{L}k\delta (k)+
\frac{1}{2L^{2}}\frac{\partial}{\partial k} \left[ k
\sum_{\pm}\Phi_{\pm}^{2}(\varphi ,k)\right]
+{\large O}\left(\frac{1}{L^{3}}\right) \right\} \right| _{k=k_{n}}
\label{7}
\end{equation}
Here the first term is the ground state energy in the absence of
the potential and flux (it is proportional to the volume: $\sim L$).
The second term ($\sim 1$) is the energy difference due the scattering
potential (in agreement with Fumi's theorem\cite{FT}), which is
flux independent. Thus, the effect of the flux is of order
$~1/L$ and is given by:
\begin{eqnarray}
\Delta E_{0}(\varphi )&=&E_{0}(\varphi )-E_{0}(0)=
\left. \sum_{n}\left\{
\frac{1}{2L^{2}}\frac{\partial}{\partial k} \left[ k
\sum_{\pm}\Phi_{\pm}^{2}(\varphi ,k)-
\Phi_{\pm}^{2}(0,k)\right]
+{\large O}\left(\frac{1}{L^{3}}\right) \right\}
\right| _{k=k_{n}} = \nonumber \\
&=&\frac{v_{_F}}{2\pi L} \left\{
F^{2}(p,\varphi )-F^{2}(p,0)\right\}
+{\large O}\left(\frac{1}{L^{2}}\right)\;.
\label{8}
\end{eqnarray}
For $N=even$ one additional particle on top of the spectrum
contributes to Eq.(\ref{7}) the term:
\begin{equation}
\Delta E_{0}^{par}(\varphi )=-{v_{F} \over L} \left\{
F(p,\varphi )-F(p,0) \right\}
\label{9}
\end{equation}
In fact, each particle contributes to the flux dependence of the
energy a term $\sim 1/L$, but the contributions of the particles with
quantum numbers $(n,+)$ and $(n,-)$, which would correspond to the
momenta $+p$ and $-p$ for $V=0$, almost cancel each other,
and the entire contribution of the Fermi sea is again of the
order of $1/L$ and converges actually just at the Fermi surface
(Eq.(\ref{8})).
For even $N$, the particle on the top, i.e.
in the state $(N/2,-)$, does not have a partner in the state
$(N/2,+)$, so the contribution of this single particle is $~1/L$;
that gives rise to the parity effect (Eq.(\ref{9})).

The formula Eq.(1) follows from Eqs.(\ref{8},\ref{9})
by a standard definition of the current $j(\varphi )=-\partial
E_{0}(\varphi )/\partial \varphi$.
The diffusion coefficient $D=-\frac{1}{2}(\partial ^{2}
E_{0}(\varphi )/\partial ^{2} \varphi )|_{\varphi =0}$
is equal to $D_{0}$ for odd $N$ and to $D_{0}+D^{par}$ for even $N$,
where
\begin{equation}
D_{0}=\frac{v_{_F}}{2 \pi L}
\sqrt{\frac{{\em T}_{F}}{1-{\em T}_{F}}}
\tan ^{-1} \sqrt{\frac{1-{\em T}_{F}}{{\em T}_{F}}};\;\;\;
 D^{par}=-\frac{v_{_F}}{2L}
\sqrt{\frac{{\em T}_{F}}{1-{\em T}_{F}}} .
\label{10}
\end{equation}
Thus, for the case of spinless fermions,
the ground state is always diamagnetic for odd $N$
and paramagnetic for even $N$.
 In the absence of  spin-orbit coupling
the spin effect can be straightforwardly taken into account.
The effect of the Fermi sea, i.e.
the quantities $j_{0}(\varphi )$, $\Delta E_{0} (\varphi )$ and
$D_{0}$, should be just multiplied by the factor $2$.
The parity effect, however, changes a little bit: the diamagnetic
ground state ($D>0$)
 is observed  only if $N=4k+2$.
So, the definite conclusion concerning a dia- or paramagnetic
nature of the ground state can be indeed drawn from the  parity
arguments only, even in the presence of a disorder potential\cite{LC}.

The transfer matrix method has already been used in the theory of
persistent current\cite{TM}.
However, it is a combination of the transfer matrix method and
the $1/L$ - expansion, which leads to the new results presented above.
For a single point scatterer $V(x)=\varepsilon\delta (x)$ the
transmission coefficient is ${\em T}_{F}=p_{F}^{2}/(p_{F}^{2}+
m^{2}\varepsilon^{2})$. Expanding our formula Eq.(1)
either in the parameter $m\varepsilon /p_{F}$ (weak potential) or
in the parameter $p_{F}/m\varepsilon$ (strong potential) we
reproduce the results for these limiting cases, which were
obtained previously\cite{TM}.
The condition $a<<L$ assumed above for simplicity is unimportant.
Actually, in the case of $a\sim L$
the derivation is essentially the same,
except of the modification due to the fact that the
forward scattering phase $\delta (p)$ in Eq.(\ref{5}) is now $\sim L$.
So, one should redefine the momentum according to
$pL=2\pi n+\delta (p)$ first and only then expand
the effect of the flux in $1/L$:
the results Eq.(1,\ref{8}-\ref{10}) will not change at all.
Simple algebra also shows that the results are valid
for arbitrary dispersion law $\varepsilon_{0}(p)$ (${\em T}_{F}$
is then understood to be defined accordingly to the
dispersion law).
The only important condition for the potential $V(x)$, which was
in fact assumed in the above derivation, is the requirement for the
transmission coefficient ${\em T}_{F}(p)$ to be a smooth function
of the momentum on the scale $1/L$ (otherwise $d{\em T}_{F}(p)/dp$
becomes $\sim L$ and the $1/L$ - expansion obviously breaks down).
Thus Eq.(1) works for the case of arbitrary single scatterer
(or several ones) and, even  under strong localization conditions
 (mean free path $l\ll L$),
 gives a correct order of magnitude for the current\cite{TM} and
predicts also a $\sim\sin\varphi$ shape of its flux dependence,
which was observed in numerical simulations for strong disorder
\cite{NR}.

In what follows we consider the effects due to electron-electron
interactions (we consider the case of a single scatterer; note however,
that in the long wave-length limit $q\to 0$ corresponding to
the Luttinger Liquid
fixed point we need only the disorder potential to be confined within
the distance $1/q$).
 We make use of Haldane's \cite{Haldane81} concept of topological
excitations in the LL ground state in the boson
representation. Due to the linear dispersion law near the Fermi points
bosonic excitations do not contribute to the current, and $j$ is
entirely defined by the topological  number $J$ which describes
the difference between the number of right- and left-moving electrons
\begin{equation}
j=-{\partial E_0 \over \partial \varphi }= {v_{_J} \over L} (J-{\varphi \over
\pi})
\;,
\label{12}
\end{equation}
where $E_0(\varphi )$ is the ground state energy of the Hamiltonian
\cite{Haldane81}
\begin{equation}
H={\pi v_{_J} \over 2L} (J-{\varphi \over \pi})^2 \;.
\label{13}
\end{equation}
In this formulation the problem is a classical one: to find the minimum of
the potential energy Eq.(\ref{13}) with an integer $J$ satisfying the parity
condition $(-1)^J=(-1)^{N+1}$.

In the boson representation the backscattering Hamiltonian has a form
\begin{equation}
H_{bsc}= V_b \sum_{J=-\infty }^{+\infty} \left(
a_{_{J+2}}^+a_{_J}\;e^{i{\hat{\theta}}} +a_{_{J-2}}^+a_{_J}\;
e^{-i{\hat{\theta}}} \right)
, \;\;\;\;\;
{\hat{\theta}}=\sqrt{g} \sum_q \sqrt{2\pi \over L\mid q \mid }
(b_q^+ + b_q) \;,
\label{14}
\end{equation}
where we introduce $a_{_J}^+$ - the creation operator of the quantum number
$J$;
$b_q^+ $ is the boson creation operator, and
 $g$ is the dimensionless electron-electron coupling constant ($g=1$ for the
 noninteracting system; $g>1$ for the attractive interaction, and $g<1$
for the case of repulsion).

Every backscattering event changes the
topological number $J$ by $\pm 2$ with a simultaneous excitation of the
bosonic environment. Now $J$ is no longer a conserved quantity but has
a tendency toward delocalization into a finite-size band. As we show
below, the delocalization of $J$ over the scale $<J^2>=R>>1$
leads to an exponential suppression of the current $\ln j\sim -R$.
One might consider
the effect of delocalization as equivalent to heating the perfect
system up to the temperature $T\sim R\: 2\pi v_{_J}/L $.
The crucial point is that the parabolic potential Eq.(\ref{13}) is very
weak (of order of $1/L$) as compared with the bare "hopping" rate $V_b$.
If the dissipation effects are neglected ($g \to 0$) then in the
ground state one finds the result $R\sim \sqrt{L}$.  However this is
never the case
because bosonic environment plays the dominant role in the formation
of the ground state.

The Hamiltonian Eq.(\ref{14}) is well known in the theory of
quantum coherence for  so-called Ohmic dissipative environments
(see, e.g. \cite{Leggett}). The effective bandwidth (or kinetic energy)
corresponding to Eq.(\ref{14}) is defined from the self-consistent equation
\begin{equation}
\Delta \approx V_b \left( {\Delta , \omega_{min} \over E_{_F} } \right) ^g \;,
\label{15}
\end{equation}
where $E_{_F}$ is the Fermi energy, and $\omega_{min} \approx
2\pi v_{_S}/L$ is the
minimal boson energy. In the simplest case
$g>1$ we immediately find that the kinetic energy is proportional to
$\Delta \sim 1/L^g$ and can be neglected as compared with the potential
term (\ref{13}). As expected, we recover the known result that backscattering
is irrelevant for the case of attractive interaction between the electrons,
in this picture, due to localization of $J$.

For $g=1$  it follows from (\ref{15}) that
$\Delta $ is proportional to $1/L$. Again, J is localized (in a sense
that fluctuations of $J$ are of order 1), but this time the solution
depends on the details of the potential, electron energy spectrum etc. in
agreement with the above discussion for the noninteracting electrons.
\documentstyle[preprint,aps,version2]{revtex}

\renewcommand{\baselinestretch}{1.5}
\renewcommand{\arraystretch}{1.5}
\setlength{\textheight}{22cm}
\setlength{\textwidth}{15.5cm}
\setlength{\evensidemargin}{0.5cm}
\setlength{\oddsidemargin}{0.5cm}
\addtolength{\topmargin}{-1cm}

\begin{document}

\vspace{3mm}
\renewcommand{\baselinestretch}{2}
\small \normalsize
\hspace{6cm} Subm. to Phys.Rev.Lett.; November 9, 1993

\begin{center}
 { \large A Simple Formula for the Persistent Current in\\
       Disordered 1D Rings: Parity and Interaction Effects}\\
 \end{center}
\vspace{3mm}
\begin{center}
 {\bf Alexander O.Gogolin$^{\dag}$ }\\
 {\em Landau Institute for Theoretical Physics,
  Russian Academy of  Sciences,} \\
 { \em Kosygina str.2, Moscow, 117940, Russia}\\
{\bf Nikolai V. Prokof'ev$^{\dag \dag }$}\\
{\em Physics Department, University of
 British Columbia, 6224 agricultural RD.,} \\
{\em Vancouver, BC, Canada V6T 1Z1}\\
\end{center}
\vspace{2mm}

\renewcommand{\baselinestretch}{2}
\small \normalsize

\begin{abstract}
The general expression for the persistent current of 1D noninteracting
electrons in a disorder potential with smooth scattering data is derived
for zero temperature.
On the basis of this expression the parity effects are discussed.
It is shown that the electron-electron interaction
may give rise to an unusual size-dependence of the current amplitude
due to  renormalized backscattering from the disorder potential.
\end{abstract}

\begin{center}
{$^{\dag}$ present address: Institute Laue-Langevin, B.P.156,
  38042 Grenoble, Cedex 9, France; Email: gogolin@gaston.ill.fr}\\
{$^{\dag \dag }$ permanent address:  Kurchatov Institute,
 Moscow, 123182, Russia}
\end{center}

\newpage

The persistent current in 1D mesoscopic rings,
being a manifestation of the Aharonov-Bohm effect for many-body systems,
addresses physical questions of principle.
Although such an effect was predicted several decades ago\cite{PP},
in normal state systems it was only recently discovered experimentally.
The experiments were performed on ensembles of many mesoscopic rings\cite{MRE}
as well as on single (or several) rings \cite{SRE}.
Generally, the theory predicts a magnitude of the current, which is much
less then one experimentally observed\cite{RT},
although there are some cases for which the theory and experiment
seem to agree\cite{NR}.
To the best of our knowledge there are no such studies  of the
interplay between the  disorder and electron-electron interaction in
1D which would
take into account nonperturbative effects in the persistent current
(on the other hand, several numerical
investigations have been curried out\cite{NR}).

The aim of this paper is to derive a simple formula for the persistent
current of noninteracting electrons in a single-channel mesoscopic ring
with a disorder potential and  to study the interplay between the
disorder and electron-electron interaction in 1D.

For $N$ spinless electrons on the ring of the length $L$ under
Aharonov-Bohm conditions the persistent current is given by the formula:
\begin{eqnarray}
&&j(\varphi )=j_{0}(\varphi )+j^{par}(\varphi )+{\large O}(1/L^{2});
\nonumber \\
&&j_{0}(\varphi )=\frac{v_{_F}}{\pi L}
\frac{\sqrt{{\em T}_{F}}\sin \varphi}{\sqrt{1-{\em T}_{F} \cos^{2} \varphi}}
F(p_{F},\varphi ); \;\;\; F(p,\varphi )=
\cos^{-1} \left( \sqrt{{\em T}(p)}\cos \varphi \right) , \\
&&j^{par}(\varphi )=-\frac{v_{_F}}{L}
\frac{\sqrt{{\em T}_{F}}\sin \varphi}{\sqrt{1-{\em T}_{F} \cos^{2} \varphi}},
\;\;\; (N=even); \;\;\;\;\;\;\; j^{par}(\varphi )=0,
\;\;\; (N= odd)\;,\nonumber
\end{eqnarray}
where $\varphi =2\pi\Phi /\Phi_{0}$ ($\Phi$ is the magnetic flux
passing through the ring) and ${\em T_{_F}}\equiv {\em T}(p{_F}) $
 is the transmission
coefficient of the ring at the Fermi energy.
The above formula explicitly shows that the persistent current
is a purely Fermi surface effect.
Moreover, it is parametrized by a single number - ${\em T_{_F}}$ (i.e.
the only characteristic of the potential,
which determines the flux dependence of the current, is ${\em T_{_F}}$)
\cite{LF}.

{\bf \em Derivation}. The single electron Schr\"{o}dinger equation
\begin{equation}
\left\{ \varepsilon_{0} \left( -i\partial_{x}
\right) +V(x)
-\varepsilon \right\} \psi =0,
\label{2}
\end{equation}
where $V(x)$ is the disorder potential and
$\varepsilon_{0}(p)$ is the dispersion law,
should be solved under the twisted boundary condition:
$\psi (x+L)=e^{i\varphi}\psi (x)$.
It was recognized already in the pioneer papers \cite{PP}
that in this Bloch functions problem
 the whole ring plays the role of the elementary cell
and $\varphi$ - the role of the quasimomentum.
The ground state energy as a function of the flux
is given by the sum
$E_{0}(\varphi )=\sum\varepsilon_{\lambda}(\varphi )$
over $N$ lowest values of the band spectrum at fixed $\varphi$.

Suppose, for clarity, that the potential $V(x)$ is localized
within the region of a radius $a<L$
(actually, the relation between $a$ and $L$ will be shown
to be unimportant).
The wave-function $\psi (x)$ can be then written in the form
$\psi (x)=Ae^{ipx}+Be^{-ipx}$ from the left of the potential
and in the form $\psi (x)=Ce^{ipx}+De^{-ipx}$ from the right of it.
The coefficients $C$ and $D$ can be expressed through $A$ and $B$
by making use of the scattering data of $V(x)$ (transfer matrix).
The twisted boundary condition  gives then
(after elementary algebra) the equations for the spectrum:
\begin{equation}
pL =  \Phi_{+}(p,\varphi ) \;\;\;( n=0);\;\;\;\;\;
pL = 2\pi n + \Phi_{\pm}(p,\varphi )\;\;\; ( n=1,2,...);  \\
\label{5}
\end{equation}
where $\Phi_{\pm}(p,\varphi )=\delta (p) \pm F(p,\varphi )$ and
$\delta (p)$ is the forward scattering phase
defined by the
scattering solution: if $A=1$ and $D=0$, then $C=\sqrt{{\em T} (p)}
e^{i\delta (p)}$.

To make progress with the Eq.(\ref{5}), the idea is to expand the
solution in $1/L$.
This is simplest when $a<<L$ (e.g. $V(x)$ represents a single
scatterer) and $\varepsilon_{0}(p)=p^{2}/2m$. Then one can write:
\begin{equation}
p_{n}=k_{n}+\frac{1}{L} \Phi_{\pm}(k_{n})+
\frac{1}{L^{2}} \Phi_{\pm}(k_{n})
\frac{\partial\Phi_{\pm}(k_{n})}{\partial k}
+{\large O}\left( \frac{1}{L^{3}}\right) ;\;\;\; \; k_{n}=2\pi n/L
\label{6}
\end{equation}
For the ground state energy we therefore have:
\begin{equation}
\left.
E_{0}=\frac{1}{m}\sum_{n}\left\{ k^{2}+\frac{2}{L}k\delta (k)+
\frac{1}{2L^{2}}\frac{\partial}{\partial k} \left[ k
\sum_{\pm}\Phi_{\pm}^{2}(\varphi ,k)\right]
+{\large O}\left(\frac{1}{L^{3}}\right) \right\} \right| _{k=k_{n}}
\label{7}
\end{equation}
Here the first term is the ground state energy in the absence of
the potential and flux (it is proportional to the volume: $\sim L$).
The second term ($\sim 1$) is the energy difference due the scattering
potential (in agreement with Fumi's theorem\cite{FT}), which is
flux independent. Thus, the effect of the flux is of order
$~1/L$ and is given by:
\begin{eqnarray}
\Delta E_{0}(\varphi )&=&E_{0}(\varphi )-E_{0}(0)=
\left. \sum_{n}\left\{
\frac{1}{2L^{2}}\frac{\partial}{\partial k} \left[ k
\sum_{\pm}\Phi_{\pm}^{2}(\varphi ,k)-
\Phi_{\pm}^{2}(0,k)\right]
+{\large O}\left(\frac{1}{L^{3}}\right) \right\}
\right| _{k=k_{n}} = \nonumber \\
&=&\frac{v_{_F}}{2\pi L} \left\{
F^{2}(p,\varphi )-F^{2}(p,0)\right\}
+{\large O}\left(\frac{1}{L^{2}}\right)\;.
\label{8}
\end{eqnarray}
For $N=even$ one additional particle on top of the spectrum
contributes to Eq.(\ref{7}) the term:
\begin{equation}
\Delta E_{0}^{par}(\varphi )=-{v_{F} \over L} \left\{
F(p,\varphi )-F(p,0) \right\}
\label{9}
\end{equation}
In fact, each particle contributes to the flux dependence of the
energy a term $\sim 1/L$, but the contributions of the particles with
quantum numbers $(n,+)$ and $(n,-)$, which would correspond to the
momenta $+p$ and $-p$ for $V=0$, almost cancel each other,
and the entire contribution of the Fermi sea is again of the
order of $1/L$ and converges actually just at the Fermi surface
(Eq.(\ref{8})).
For even $N$, the particle on the top, i.e.
in the state $(N/2,-)$, does not have a partner in the state
$(N/2,+)$, so the contribution of this single particle is $~1/L$;
that gives rise to the parity effect (Eq.(\ref{9})).

The formula Eq.(1) follows from Eqs.(\ref{8},\ref{9})
by a standard definition of the current $j(\varphi )=-\partial
E_{0}(\varphi )/\partial \varphi$.
The diffusion coefficient $D=-\frac{1}{2}(\partial ^{2}
E_{0}(\varphi )/\partial ^{2} \varphi )|_{\varphi =0}$
is equal to $D_{0}$ for odd $N$ and to $D_{0}+D^{par}$ for even $N$,
where
\begin{equation}
D_{0}=\frac{v_{_F}}{2 \pi L}
\sqrt{\frac{{\em T}_{F}}{1-{\em T}_{F}}}
\tan ^{-1} \sqrt{\frac{1-{\em T}_{F}}{{\em T}_{F}}};\;\;\;
 D^{par}=-\frac{v_{_F}}{2L}
\sqrt{\frac{{\em T}_{F}}{1-{\em T}_{F}}} .
\label{10}
\end{equation}
Thus, for the case of spinless fermions,
the ground state is always diamagnetic for odd $N$
and paramagnetic for even $N$.
 In the absence of  spin-orbit coupling
the spin effect can be straightforwardly taken into account.
The effect of the Fermi sea, i.e.
the quantities $j_{0}(\varphi )$, $\Delta E_{0} (\varphi )$ and
$D_{0}$, should be just multiplied by the factor $2$.
The parity effect, however, changes a little bit: the diamagnetic
ground state ($D>0$)
 is observed  only if $N=4k+2$.
So, the definite conclusion concerning a dia- or paramagnetic
nature of the ground state can be indeed drawn from the  parity
arguments only, even in the presence of a disorder potential\cite{LC}.

The transfer matrix method has already been used in the theory of
persistent current\cite{TM}.
However, it is a combination of the transfer matrix method and
the $1/L$ - expansion, which leads to the new results presented above.
For a single point scatterer $V(x)=\varepsilon\delta (x)$ the
transmission coefficient is ${\em T}_{F}=p_{F}^{2}/(p_{F}^{2}+
m^{2}\varepsilon^{2})$. Expanding our formula Eq.(1)
either in the parameter $m\varepsilon /p_{F}$ (weak potential) or
in the parameter $p_{F}/m\varepsilon$ (strong potential) we
reproduce the results for these limiting cases, which were
obtained previously\cite{TM}.
The condition $a<<L$ assumed above for simplicity is unimportant.
Actually, in the case of $a\sim L$
the derivation is essentially the same,
except of the modification due to the fact that the
forward scattering phase $\delta (p)$ in Eq.(\ref{5}) is now $\sim L$.
So, one should redefine the momentum according to
$pL=2\pi n+\delta (p)$ first and only then expand
the effect of the flux in $1/L$:
the results Eq.(1,\ref{8}-\ref{10}) will not change at all.
Simple algebra also shows that the results are valid
for arbitrary dispersion law $\varepsilon_{0}(p)$ (${\em T}_{F}$
is then understood to be defined accordingly to the
dispersion law).
The only important condition for the potential $V(x)$, which was
in fact assumed in the above derivation, is the requirement for the
transmission coefficient ${\em T}_{F}(p)$ to be a smooth function
of the momentum on the scale $1/L$ (otherwise $d{\em T}_{F}(p)/dp$
becomes $\sim L$ and the $1/L$ - expansion obviously breaks down).
Thus Eq.(1) works for the case of arbitrary single scatterer
(or several ones) and, even  under strong localization conditions
 (mean free path $l\ll L$),
 gives a correct order of magnitude for the current\cite{TM} and
predicts also a $\sim\sin\varphi$ shape of its flux dependence,
which was observed in numerical simulations for strong disorder
\cite{NR}.

In what follows we consider the effects due to electron-electron
interactions (we consider the case of a single scatterer; note however,
that in the long wave-length limit $q\to 0$ corresponding to
the Luttinger Liquid
fixed point we need only the disorder potential to be confined within
the distance $1/q$).
 We make use of Haldane's \cite{Haldane81} concept of topological
excitations in the LL ground state in the boson
representation. Due to the linear dispersion law near the Fermi points
bosonic excitations do not contribute to the current, and $j$ is
entirely defined by the topological  number $J$ which describes
the difference between the number of right- and left-moving electrons
\begin{equation}
j=-{\partial E_0 \over \partial \varphi }= {v_{_J} \over L} (J-{\varphi \over
\pi})
\;,
\label{12}
\end{equation}
where $E_0(\varphi )$ is the ground state energy of the Hamiltonian
\cite{Haldane81}
\begin{equation}
H={\pi v_{_J} \over 2L} (J-{\varphi \over \pi})^2 \;.
\label{13}
\end{equation}
In this formulation the problem is a classical one: to find the minimum of
the potential energy Eq.(\ref{13}) with an integer $J$ satisfying the parity
condition $(-1)^J=(-1)^{N+1}$.

In the boson representation the backscattering Hamiltonian has a form
\begin{equation}
H_{bsc}= V_b \sum_{J=-\infty }^{+\infty} \left(
a_{_{J+2}}^+a_{_J}\;e^{i{\hat{\theta}}} +a_{_{J-2}}^+a_{_J}\;
e^{-i{\hat{\theta}}} \right)
, \;\;\;\;\;
{\hat{\theta}}=\sqrt{g} \sum_q \sqrt{2\pi \over L\mid q \mid }
(b_q^+ + b_q) \;,
\label{14}
\end{equation}
where we introduce $a_{_J}^+$ - the creation operator of the quantum number
$J$;
$b_q^+ $ is the boson creation operator, and
 $g$ is the dimensionless electron-electron coupling constant ($g=1$ for the
 noninteracting system; $g>1$ for the attractive interaction, and $g<1$
for the case of repulsion).

Every backscattering event changes the
topological number $J$ by $\pm 2$ with a simultaneous excitation of the
bosonic environment. Now $J$ is no longer a conserved quantity but has
a tendency toward delocalization into a finite-size band. As we show
below, the delocalization of $J$ over the scale $<J^2>=R>>1$
leads to an exponential suppression of the current $\ln j\sim -R$.
One might consider
the effect of delocalization as equivalent to heating the perfect
system up to the temperature $T\sim R\: 2\pi v_{_J}/L $.
The crucial point is that the parabolic potential Eq.(\ref{13}) is very
weak (of order of $1/L$) as compared with the bare "hopping" rate $V_b$.
If the dissipation effects are neglected ($g \to 0$) then in the
ground state one finds the result $R\sim \sqrt{L}$.  However this is
never the case
because bosonic environment plays the dominant role in the formation
of the ground state.

The Hamiltonian Eq.(\ref{14}) is well known in the theory of
quantum coherence for  so-called Ohmic dissipative environments
(see, e.g. \cite{Leggett}). The effective bandwidth (or kinetic energy)
corresponding to Eq.(\ref{14}) is defined from the self-consistent equation
\begin{equation}
\Delta \approx V_b \left( {\Delta , \omega_{min} \over E_{_F} } \right) ^g \;,
\label{15}
\end{equation}
where $E_{_F}$ is the Fermi energy, and $\omega_{min} \approx
2\pi v_{_S}/L$ is the
minimal boson energy. In the simplest case
$g>1$ we immediately find that the kinetic energy is proportional to
$\Delta \sim 1/L^g$ and can be neglected as compared with the potential
term (\ref{13}). As expected, we recover the known result that backscattering
is irrelevant for the case of attractive interaction between the electrons,
in this picture, due to localization of $J$.

For $g=1$  it follows from (\ref{15}) that
$\Delta $ is proportional to $1/L$. Again, J is localized (in a sense
that fluctuations of $J$ are of order 1), but this time the solution
depends on the details of the potential, electron energy spectrum etc. in
agreement with the above discussion for the noninteracting electrons.

In the most intriguing case $g<1$ the coherence is restored between the
neighbouring values of $J$ at frequencies below
\begin{equation}
\Delta \approx E_{_F} \left( {V_b \over E_{_F}} \right) ^{1/(1-g)} \;,
\label{16}
\end{equation}
with delocalization of $J$ in the ground state.
Let us start with the extreme case of $g=0$.
In the "momentum" representation the Hamiltonian (\ref{13})-(\ref{14})
can be conveniently rewritten in the form (one might identify the momentum $p$
with the canonical conjugate phase field $\overline{\theta }_{J} $;
$[J,\overline{\theta }_{J}]=i$, see Ref.\cite{Haldane81}):
\begin{equation}
H=-{2\pi v_{_J} \over L} {\partial ^2 \over \partial  p^2} + 2\Delta \cos p \;,
\label{17}
\end{equation}
with twisted periodic condition $\psi (p+2\pi )=e^{i\varphi } \psi (p)$ (we
assumed $J=even$ here; the case of $J=odd$ will correspond to $\varphi \to
\varphi + \pi $). Since the effective mass in the momentum representation
is proportional to $1/L$ we can consider this Hamiltonian as a standard
tight-binding model with the energy of the lowest state being defined as
$E_0(\varphi )=const + D(1-\cos \varphi )$,
and the amplitude $D$ being calculated in the semiclassical approach
\begin{equation}
D\approx \sqrt{8v_{_J}\Delta \over \pi L} \exp \left(
-8 \sqrt{L\Delta \over 2\pi v_{_J}} \right) \;.
\label{19}
\end{equation}
Thus, for the case $g=0$, the persistent current,
$j\approx D(-1)^N \sin \varphi$,
 would be exponentially suppressed.

With $g>0$, however, the concept of the effective mass for $J$
completely fails even in the ground state. The dynamics of $J$ is
overdamped as follows from the constant mobility in the zero-temperature
limit \cite{Schmid}, $\mu =1/(2\pi g)$. In the presence of the
parabolic potential Eq.(\ref{13}) the delocalization of $J$ in the ground state
depends on $L$ only logarithmically \cite{p},
$ <J^2>={\mu \over \pi }\ln \left( {L\Delta  / (2\pi v_{_J} \mu^2} )\right) $,
and one might expect a power dependence $J(L)$. It is easy to show \cite{tobe},
that in the momentum representation the problem can be reduced to the
study of the effective action $\int {\it D}p \exp (-S_{eff})$
\begin{equation}
S_{eff}=\int_{-\beta /2}^{\beta /2} d\tau \;2\Delta \cos p(\tau )
+\beta \sum_n {\omega_n^2 \mid p_n \mid^2 \over 2(4\pi v_{_J}/L +2\pi g
\mid \omega_n \mid )} \;;\;\;\;\; (\beta \to \infty) \;,
\label{22}
\end{equation}
where $\omega_n=2\pi n/\beta $, and $p_n$ is the Fourie-transform of
$p(\tau )$. Except very low frequencies $\omega_n \le 2v_{_J}/gL$ the
effective action is governed by the dissipative term ${\beta \over 4\pi g }
\sum \mid \omega \mid \mid p_\omega \mid ^2 $. The  quasiclassical solution
for the overdamped motion in the $\cos $-barrier was
studied in Ref.\cite{Korsh}, so we can readily write down the expression
 for the current amplitude $D$  in the leading logarithmic approximation
\begin{equation}
D\sim g\Delta  \exp \left(
-{1 \over g} \ln \bigg[ { L\Delta \over 2\pi v_{_J} }\bigg] \right) \;.
\label{23}
\end{equation}
Thus our basic result for the persistent current in the case of
repulsion between the electrons has a form
\begin{equation}
j\sim g\Delta   \left(
 2\pi v_{_J}\over  L\Delta \right) ^{1/g} (-1)^N \sin \varphi \;.
\label{24}
\end{equation}

The suppression of the current to higher order in $1/L$ is closely
related to the result of Ref.\cite{Kane} that in the limit
$E \to E_{_F}$ backscattering renormalizes into a perfect reflection
from the potential. Effectively the transmission coefficient near
the Fermi points has a power law dependence \cite{Kane}
\begin{equation}
T(E) \sim T_0 \left( {E-E_{_F} \over \Delta }\right) ^{2(1/g-1)} \;,
\label{25}
\end{equation}
leading to $T_{_F} \sim T_0 L^{-2(1/g-1)}$. If we substitute this result
into the formula Eq.(\ref{1}), then in the limit $T_{_F}\ll 1$ we easily
obtain the same scaling behavior Eq.(\ref{24}) \cite{comment}.

To summarize, we derive for the first time an explicit analytic solution for
the persistent current of noninteracting electrons in disordered
1D ring under quite general assumptions about the disorder potential.
We find that the transmission coefficient of the ring at the Fermi energy
is the only relevant parameter in the problem. The case of spinless
electrons is solved including electron-electron interactions, and
a nontrivial size dependence
of the persistent current with interaction-dependent exponent is obtained
(provided the parameter $\Delta $, Eq. (\ref{16}), is large enough
compared with $ v_{_J}2\pi /L$).
It seems interesting to generalize the above approach to
the case of  spin-orbit coupling and to the (more realistic)
case of multi-channel rings for which the existence of a
formula expressing the persistent current in terms of the
transmittance matrix at the Fermi energy is still expected
(although the flux dependence of the current will be more
complicated). These generalizations are now in progress \cite{tobe}.

We are thankful to
P.C.E.Stamp,
A.G.Aronov, H.Cappelmann, A.Ioselevich, A.I.Larkin and M.Fabrizio
for valuable discussions.
One of us (N.V.P.) was supported by NSERC.


\newpage

\begin{thebibliography}{99}

\bibitem{PP}
   N.Byers and C.N.Yang, Phys.Rev.Lett.{\bf 7}, 46(1961);
   W.Kohn, Phys.Rev. {\bf 133}, A171(1964);
   F.Bloch, Phys.Rev. {\bf 21}, 1241(1968);
   L.Gunther and Y.Imry, Solid State Commun.{\bf 7} 1394(1969);
   M.B\"{u}ttiker, Y.Imry and R.Landauer, Phys.Lett.{\bf 96A}, 365(1983).

\bibitem{MRE}
   L.P.Levy,G.Dolan, J.Dunsmuir and H.Brouchiat,
   Phys.Rev.Lett. {\bf 64}, 2074 (1990).

\bibitem{SRE}
   V.Chandrasekhar, R.A.Webb, M.J.Brady, M.B.Ketchen, W.J.Galager and
   A.Kleinsasser,
   Phys.Rev.Lett. {\bf 67}, 3578 (1991).

\bibitem{RT}
   Recently the literature on the persistent current problem
   has been really exploding.
   For recent perturbation theory calculations see
   R.A.Smith and V.Ambegaokar, Europhys.Lett. {\bf 20}, 161(1992)
   and references therein.

\bibitem{NR}
   M.Abraham and R.Berkovits, Rhys.Rev.Lett. {\bf 70}, 1509(1993);
   G.Bouzerar, D.Poilblanc and G.Montambaux,
   "Persistent Current in 1D Disordered Rings of
   Interacting Electrons",
   preprint(1993).

\bibitem{LF}
   This formula can be considered as an analog of the
   famous Landauer formula for the conductance;
   R.Landauer, IBM J.Res.Dev. {\bf 1}, 223(1957).

\bibitem{FT}
   F.G.Fumi, Philos.Mag. {\bf 46}, 1007(1955).

\bibitem{LC}
   That may be considered as a
   proof of the so-called Leggett conjecture;
   A.J.Leggett, in {\em Granular Nanoelectronics}, edited by
   D.K.Ferry, J.R.Barker and C.Jacoboni, NATO ASI Ser. {\bf B251}
   (Plenum, New York, 1991), p.297;
   see also D.Loss, Phys.Rev.Lett. {\bf 69}, 343(1992).

\bibitem{TM}
   H.F.Cheung, Y.Gefen, E.K.Riedel and W.H.Shih,
   Phys.Rev.{\bf B37}, 6050(1988).

\bibitem{Haldane81}
     F.D.M.Haldane, J.Phys.{\bf C}, {\bf 14}, 2585(1981).

\bibitem{Kane}
C.L.Kane  and M.P.A.Fisher, Phys.Rev.Lett.
{\bf 68}, 1220(1992);
 K.A.Matveev, D.Yue and L.I.Glazman,
 "Conduction of a Weakly Interacting 1D Electron Gas through a
 Single Barrier", preprint(1993).

\bibitem{Leggett}
 A.J.Leggett, S.Chakravarty, A.T.Dorsey,
M.P.A.Fisher, A.Garg and W.Zwerger, Rev.Mod.Phys.{\bf 59}, 1(1987).


\bibitem{Schmid}
 A.Schmid, Phys.Rev.Lett.{\bf 51}, 1506(1983);
S.A.Bulgadaev, Zh.Teor.Eksp.Fiz.{\bf 90}, 634(1986) [Sov.Phys.JETP {\bf 63},
369(1986)].

\bibitem{p}
N.V.Prokof'ev, Int.J.Mod.Phys.{\bf B 7}, 3327(1993).

\bibitem{tobe}
 A.O.Gogolin, N.V.Prokof'ev, to be published.

\bibitem{Korsh}
S.E.Korshunov, Zh.Eksp.Teor.Fiz.{\bf 92}, 1828(1987) [Sov.Phys.JETP
{\bf 65}, 1025(1987)].

\bibitem{comment}
   In fact, such a substitution turns out to be correct for
   all such "Fermi surface" effects: e.g.,
   the conductance (Ref.\cite{Kane}); the
   X-Ray problem (A.O.Gogolin, Phys.Rev.Lett.{\bf 71}, 2995(1993);
   N.V.Prokof'ev, Phys.Rev.{\bf B} (1993), in press) and, now, the
   persistent current.

\end{thebibliography}
\end{document}